\documentclass[conference,10pt]{IEEEtran}
\usepackage[T1]{fontenc}
\usepackage[latin9]{inputenc}
\usepackage{color}
\usepackage{float}
\usepackage{amsthm}
\usepackage[nosumlimits]{amsmath}
\usepackage{amssymb}
\usepackage{multirow}

\usepackage{hyperref}
\usepackage{breakurl}
{
\makeatother

\floatstyle{ruled}
\newfloat{algorithm}{tbp}{loa}
\providecommand{\algorithmname}{Algorithm}
\floatname{algorithm}{\protect\algorithmname}

\theoremstyle{plain}

\theoremstyle{definition}

\theoremstyle{plain}

\theoremstyle{remark}

\theoremstyle{remark}

\allowdisplaybreaks

\usepackage{mathtools}

\usepackage{amsfonts}
\usepackage{cite}
\usepackage{array}
\usepackage{algorithm}
\usepackage{algorithmic}
\usepackage{graphicx}
\usepackage{subfigure}

\DeclareMathOperator*{\maxi}{max}

\DeclareMathOperator*{\st}{s.t.}

\newcommand{\herm}{^{H}}

\setlength{\textfloatsep}{5pt}

\newcommand*{\rom}[1]{\expandafter\@slowromancap\romannumeral #1@}

\makeatother

\providecommand{\definitionname}{Definition}
\providecommand{\factname}{Fact}
\providecommand{\remarkname}{Remark}
\providecommand{\theoremname}{Theorem}
\providecommand{\lemmaname}{Lemma}

\mathchardef\mhyphen="2D

\IEEEoverridecommandlockouts

\begin{document}
\title{Energy-Efficient Joint Unicast and Multicast Beamforming with Multi-Antenna User Terminals}

\author{\IEEEauthorblockN{Oskari Tervo\textsuperscript{$\ast$}, Le-Nam Tran\textsuperscript{$\dag$}, Symeon Chatzinotas\textsuperscript{$\ddagger$}, Markku Juntti\textsuperscript{$\ast$}, and Bj\"orn Ottersten\textsuperscript{$\ddagger$}}
\IEEEauthorblockA{\textsuperscript{$\ast$}Centre for Wireless Communications, University of Oulu, Oulu, Finland\\
\textsuperscript{$\dag$}Department of Electronic Engineering, Maynooth University, Maynooth, Co Kildare, Ireland\\
\textsuperscript{$\ddagger$}the Interdisciplinary Centre for Security, Reliability and Trust, University of Luxembourg, Luxembourg \\
Email: \{firstname.lastname\}@oulu.fi, lenam.tran@nuim.ie, \{symeon.chatzinotas, bjorn.ottersten\}@uni.lu} 
}


\maketitle

\begin{abstract}
This paper studies energy-efficient joint transmit and receive beamforming in multi-cell multi-user multiple-input multiple-output systems. We consider conventional network energy efficiency metric where the users can receive unicasting streams in addition to the group-specific common multicasting streams which have certain rate constraints. The goal is to use the transmission resources more efficiently to improve the energy efficiency, when the users are equipped with multiple antennas. Numerical results show the achieved energy efficiency gains by using the additional degrees of freedom of the multicasting transmission to private message unicasting.

\end{abstract}


\setlength{\textfloatsep}{3pt }

\section{\label{sec:intro}Introduction}

%
Achieving high energy efficiency (EE) has become one of the main targets in the future wireless communications standards \cite{Ericsson}. To address this challenge, the cellular networks have to be based on aggressive frequency reuse. Due to the fact that different transmitters communicate in the same frequency spectrum, this approach results in high interference conditions. To achieve high performance in this case, multi-antenna techniques can be used to control the interference by forming beams in specific directions. An efficient method based on this approach is coordinated beamforming \cite{Irmer-11}, which designs all the beams in a coordinated manner. The energy-efficient coordinated beamforming methods have been widely studied in the literature, e.g., in \cite{Tervo-16Arxiv,Nguyen-15,TervoWsumEE-15}.


Future networks are facing another important challenge as the type of communications is changing. The massive popularity of smart mobile devices and the development of mobile services means that the requested data can be highly correlated between many users. For example, during a popular event or in a shopping mall, groups of people may be consuming the same multimedia content. In these cases, multicasting transmission \cite{Xiang-13,PennanenICC-16,Tervo-17,Joudeh-16} is an efficient method to improve EE, i.e., the same information can be transmitted only once to all the users requesting it \cite{Alcatel-Lucent}. Coordinated multicast beamforming has been studied, e.g., for transmit power minimization \cite{Xiang-13,PennanenICC-16}, max-min fairness \cite{Xiang-13} and energy efficiency maximization \cite{Tervo-17}.

Prior work has been mostly focusing on either unicast or multicast beamforming, and, thus, transmission resources may be potentially used inefficiently. For example, in multicasting, the achievable data rate is determined by the weakest link in the multicasting group, and, thus, the users with the strongest links can have a significant potential to use some of the resources to receive conventional unicasting streams. Conventionally, unicasting and multicasting services have been scheduled/allocated in different time or frequency resources. However, more contemporary research has been investigating more efficient and flexible use of unicasting and multicasting. For example, one option is to use so-called layered division multiplexing \cite{Zhang-16,Zhao-16,Zhang-16TandP}, where the common data is decoded first by treating private stream interference as noise, and then the successive interference cancellation (SIC) is invoked to decode the private data. This is possible even when the users are equipped with only a single antenna, but comes at a price of additional complexity at  the receivers. On the other hand, different services can be separated in the spatial domain even without SIC when the users are equipped with multiple receive antennas. By using linear processing at the receiver side, the users can receive different services utilizing the same resources to improve both spectral and energy efficiency. 

In this paper, we study energy-efficient joint transmit and receive beamforming in multi-cell multi-user multiple-input multiple-output (MIMO) systems. By treating all the interference as noise, we aim at maximizing the EE by allowing the users to receive private data streams in addition to the group-specific common data with certain rate constraints. By the equivalence between minimum mean squared error (MMSE) and signal-to-interference-plus-noise ratio (SINR), we develop a successive convex approximation (SCA) based algorithm, where the transmit beamformers, the operating parameters resulting from the linearization, and the MMSE receivers are updated sequentially. The numerical results illustrate the monotonic convergence of the algorithm and demonstrate that it is energy-efficient to use the additional transmission resources for unicasting services when the users are equipped with multiple receive antennas. 

Contrary to existing work \cite{Tervo-17}, where energy-efficient multicast coordinated transmit beamforming with single-antenna users was considered, this paper resolves the multicasting inefficiency by joint design of multicasting and unicasting. Another important contribution compared to existing works is to consider multi-antenna user terminals, so that both the transmit and receive beamformers are jointly optimized to maximize the energy efficiency.



\subsubsection*{Notations} $|x|$ denotes the cardinality of $x$ if $x$ is a set, or absolute value of $x$, otherwise. $||\mathbf{x}||_2$ is the Euclidean norm of $\mathbf{x}$ and boldcase letters are vectors. $\mathbf{x}\herm$ denotes Hermitian transpose of $\mathbf{x}$.

\section{\label{PF} System Model and Problem Formulation}
\label{sec:ProblemFormulation}
\subsection{System Model}
We consider a multi-cell system, where a set of $B=|\mathcal{B}| (\mathcal{B}=\{1,\ldots,B\})$ BSs with $N_b$ antennas at BS $b$ transmits data to $G=|\mathcal{G}| (\mathcal{G}=\{1,\ldots,G\})$ groups of users with $M_k$ antennas at user $k$. The users are divided into groups so that the users inside the same group request the same information, i.e., the data can be multicasted to a group. The set of groups served by BS $b$ is denoted as $\mathcal{G}_b$, the user set belonging to a group $g$ is denoted as $\mathcal{K}_g$, and $b_k \in \mathcal{B}$ denotes the serving BS of either user $k$ or user group $k$. The set of all users is denoted as $\mathcal{K}$. Each user belongs to only one multicasting group. The set of users served by BS $b$ is denoted as $\mathcal{U}_b$. Since the users are equipped with multiple antennas, they can also receive private data at the same resources to improve the energy efficiency. That is, the base stations can jointly unicast and multicast in the same resources. It is assumed that the multicasting information is always transmitted, i.e., user $k$ can receive up to $L_k=|\mathcal{L}_k| (\mathcal{L}_k=\{1,\ldots,M_k-1\})$ independent private data streams in addition to the multicasting information.
The received signals for the private stream $l$ and common stream at user $k$ can be written as
\begin{eqnarray} \label{eq:RxSignal}
\mathbf{y}_{k,l} &=& \overbrace{\mathbf{H}_{b_k,k} \mathbf{w}_{k,l} {s}_{k,l}}^\text{desired signal}  + \overbrace{\sum\limits_{i \in \mathcal{K}}\sum\limits_{\substack{j \in \mathcal{L}_i,\\(i,j)\neq(k,l)}} \mathbf{H}_{b_i,k} \mathbf{w}_{i,j} {s}_{i,j}}^\text{private inter-stream interference} \nonumber \\
&  & + \sum_{u\in\mathcal{G}}\overbrace{\mathbf{H}_{b_u,k} \tilde{\mathbf{w}}_{u,c} {s}_{u,c}}^\text{common stream interference} + \mathbf{n}_{k}
\end{eqnarray}
\begin{eqnarray} \label{eq:RxSignal}
\mathbf{y}_{k,c} &=& \overbrace{\mathbf{H}_{b_g,k} \tilde{\mathbf{w}}_{g,c} {s}_{g,c}}^\text{desired signal}  + \overbrace{\sum\limits_{i \in \mathcal{K}}\sum\limits_{j \in \mathcal{L}_i} \mathbf{H}_{b_i,k} \mathbf{w}_{i,j} {s}_{i,j}}^\text{private inter-stream interference} \nonumber \\
&  & + \sum_{u\in\mathcal{G}\setminus \{g\}}\overbrace{\mathbf{H}_{b_u,k} \tilde{\mathbf{w}}_{u,c} {s}_{u,c}}^\text{common stream interference} + \mathbf{n}_{k},
\end{eqnarray}
respectively, where $\mathbf{H}_{b,k} \in \mathcal{C}^{M_k \times N_b}$ is the MIMO channel matrix from BS $b$ to user $k$, $\mathbf{w}_{k,l} \in \mathcal{C}^{N_b \times 1}$ is the transmit beamforming vector of the private stream $l$ to user $k$, $\tilde{\mathbf{w}}_{g,c} \in \mathcal{C}^{N_b \times 1}$ is the transmit beamforming vector for the common stream to user group $g$, ${s}_{k,l} \in \mathcal{C}$ and ${s}_{g,c} \in \mathcal{C}$ is the normalized $l$th private and common data symbol for user $k$ and user group $g$, respectively, and $\mathbf{n}_{k} \in \mathcal{C}^{M_k \times 1} $ is the complex white Gaussian receiver noise with zero mean and variance $\sigma_{k}^{2}$ per element. User $k$ employs a receive beamformer $\mathbf{u}_{k,l} \in \mathcal{C}^{M_k\times 1}$ for the data stream $l$.
We assume that data decoding is done simultaneously and independently for unicasting and multicasting. Thus, the stream specific SINR expressions of user $k$ can be written as
\begin{equation}
\Gamma_{k,l}(\mathbf{u}_{k,l},\mathbf{w}) \triangleq \frac{|\mathbf{u}_{k,l}\mathbf{H}_{b_k,k} \mathbf{w}_{k,l}|^{2}}{{||\mathbf{u}_{k,l}||_2^2N_0 + I_{k,l}(\mathbf{u}_{k,l},\mathbf{w})}}
\end{equation}
and
\begin{equation}
\Gamma_{k,c}(\mathbf{u}_{k,c},\mathbf{w}) \triangleq \frac{|\mathbf{u}_{k,c}\mathbf{H}_{b_k,k} \tilde{\mathbf{w}}_{g,c}|^{2}}{{||\mathbf{u}_{k,c}||_2^2N_0 + I_{k,c}(\mathbf{u}_{k,c},\mathbf{w})}}
\end{equation}
for the private and common data streams, respectively, where
\begin{eqnarray}
I_{k,l}(\mathbf{u}_{k,l},\mathbf{w})\triangleq \sum\limits_{i \in \mathcal{K}}\sum\limits_{\substack{j \in \mathcal{L}_i,\\(i,j)\neq(k,l)}} |\mathbf{u}_{k,l}\mathbf{H}_{b_i,k} \mathbf{w}_{i,j}|^2 \nonumber \\ + \sum_{u\in\mathcal{G}}|\mathbf{u}_{k,l}\mathbf{H}_{b_u,k} \tilde{\mathbf{w}}_{u,c}|^2
\end{eqnarray}
and
\begin{eqnarray}
I_{k,c}(\mathbf{u}_{k,c},\mathbf{w})\triangleq \sum\limits_{i \in \mathcal{K}}\sum\limits_{j \in \mathcal{L}_i} |\mathbf{u}_{k,c}\mathbf{H}_{b_i,k} \mathbf{w}_{i,j}|^2\nonumber \\ + \sum_{u\in\mathcal{G}\setminus \{g\}}|\mathbf{u}_{k,c}\mathbf{H}_{b_u,k} \tilde{\mathbf{w}}_{u,c}|^2.
\end{eqnarray}
are the interference experienced by the $l$th private stream of user $k$ and the common stream of user $k$, respectively, and $N_0$ is the noise power over the total transmission bandwidth $W$.
The data rate of stream $l$ for user $k$ is given as
\begin{equation}
R_{k,l}(\mathbf{u}_{k,l},\mathbf{w}) \triangleq W\log(1+\Gamma_{k,l}(\mathbf{u}_{k,l},\mathbf{w}))
\end{equation}
and for the common stream, the achievable data rate for user group $g$ is determined by the worst user in the group, i.e.,
\begin{equation}
R_{g,c}(\{\mathbf{u}_{k,c}\}_{k\in\mathcal{K}_g},\mathbf{w}) \triangleq W\underset{k\in\mathcal{K}_g}{\min}\log(1+\Gamma_{k,c}(\mathbf{u}_{k,c},\mathbf{w})).
\end{equation}
It is worth observing that the common data rate of group $g$ is a function of all the receive beamformers in the group.
In the receiver, the mean squared error (MSE) expressions become
\begin{eqnarray}
\epsilon_{k,l} & \triangleq & \mathbb{E}[|\mathbf{u}_{k,l}\herm\mathbf{y}_{k,l} - s_{k,l}|^2] =  |1-\mathbf{u}_{k,l}\herm\mathbf{H}_{b_k,k}\mathbf{w}_{k,l}|^2\nonumber \\
& + & I_{k,l}(\mathbf{u}_{k,l},\mathbf{w}) +  N_0||\mathbf{u}_{k,l}||_2^2.\label{MSEprivate}
\end{eqnarray}
\begin{eqnarray}
\epsilon_{k,c} \triangleq \mathbb{E}[|\mathbf{u}_{k,c}\herm\mathbf{y}_{k,c} - s_{g,c}|^2] = |1-\mathbf{u}_{k,c}\herm\mathbf{H}_{b_k,k}\tilde{\mathbf{w}}_{g,c}|^2 \nonumber \\ + I_{k,c}(\mathbf{u}_{k,c},\mathbf{w}) + N_0||\mathbf{u}_{k,c}||_2^2.\label{MSEcommon}
\end{eqnarray}

\subsection{Power Consumption Model}

The total power consumption at the transmitter side is \cite{Tervo-15}
\begin{equation}
\label{eq:powermodel}
\begin{aligned}
P_{\text{tot}}\triangleq & \dfrac{1}{\eta}(\sum\limits_{k\in\mathcal{K}}\sum\limits_{l\in\mathcal{L}_k}||\mathbf{w}_{k,l}||_2^{2}+\sum\limits_{g\in\mathcal{G}}||\tilde{\mathbf{w}}_{g,c}||_2^{2}) \\ & + \sum\limits_{b\in\mathcal{B}}(P_{0,\text{BS}} + N_b P_{\text{RF,BS}}) \\ & + \sum\limits_{k\in\mathcal{K}}(P_{0,\text{UE}} + M_k P_{\text{RF,UE}})
\end{aligned}
\end{equation}
where $\eta\in[0,1]$ is the power amplifier
efficiency, which is assumed to be fixed for simplicity,
$P_{0,\text{BS}}$ and $P_{0,\text{UE}}$ is the fixed circuit power consumption of each BS and each user, respectively, $P_{\text{RF,BS}}$ and $P_{\text{RF,UE}}$ is the power consumption of an active radio frequency (RF) chain at BS and user, respectively. For the ease of notation, the aggregated circuit power is denoted by $P_\text{cir}\triangleq \sum\limits_{b\in\mathcal{B}}(P_{0,\text{BS}} + N_b P_{\text{RF,BS}}) + \sum\limits_{k\in\mathcal{K}}(P_{0,\text{UE}} + M_k P_{\text{RF,UE}})$.

\subsection{Problem Formulation }
We aim at maximizing the network energy efficiency, which is expressed as
\begin{subequations} \label{EEmax}
\begin{eqnarray}{} & \hspace{-15mm} \underset{\mathbf{w},\mathbf{u}}{\maxi}    & \hspace{-8mm} \frac{\sum\limits_{k\in\mathcal{K}}\sum\limits_{l\in\mathcal{L}_k}R_{k,l}(\mathbf{u}_{k,l},\mathbf{w}) + \sum\limits_{g\in\mathcal{G}}R_{g,c}(\{\mathbf{u}_{k,c}\}_{k\in\mathcal{K}_g},\mathbf{w})}{\dfrac{1}{\eta}(\sum\limits_{k\in\mathcal{K}}\sum\limits_{l\in\mathcal{L}_k}||\mathbf{w}_{k,l}||_2^{2}+\sum\limits_{g\in\mathcal{G}}||\tilde{\mathbf{w}}_{g,c}||_2^{2}) + P_\text{cir}}  \label{eq:EE:obj} \\ \hspace{-5mm} \st
 &   & \hspace{-5mm} \sum\limits_{k\in\mathcal{U}_b}\sum\limits_{l\in\mathcal{L}_k}||\mathbf{w}_{k,l}||_2^2 + \sum\limits_{g\in\mathcal{G}_b}||\tilde{\mathbf{w}}_{g,c}||_2^{2} \leq P_b, \forall b \in \mathcal{B} \label{PC} \\
 &   & \hspace{-5mm} R_{g,c}(\{\mathbf{u}_{k,c}\}_{k\in\mathcal{K}_g},\mathbf{w}) \geq \bar{R}_{g,c}, \forall g \in \mathcal{G}
\end{eqnarray}
\end{subequations}
where \eqref{PC} are BS-specific power constraints and $\bar{R}_{g,c}$ is the rate requirement for the common stream in the multicasting group $g$.\footnote{We could similarly add rate constraints for the unicasting streams. However, since this paper investigates the potential network EE gains, the unicasting fairness is not considered. Note that the problem is infeasible if the number of antennas at some base station $b$ is smaller than the number of groups served by the BS. Thus, we assume $N_b\geq |\mathcal{G}_b|$ throughout the paper.}
The above problem is a non-convex fractional program which is hard to tackle as such due to the non-convex rate functions in the objective and constraints.

\section{Proposed Solution}\label{sec:CentralizedMethods}

To find a more tractable reformulation, we first use the well-known equivalence between the SINR and MSE when the optimal MMSE receivers are used \cite{Shi:WMMSE:2011}. Specifically, we can equivalently write \eqref{EEmax} as
\begin{subequations}
\label{eq:EEmax:reform1}
\begin{eqnarray}
& \hspace{-15mm} \underset{\mathbf{w},\mathbf{u}}{\maxi} & \hspace{-2mm}   \frac{\sum\limits_{k\in\mathcal{K}}\sum\limits_{l\in\mathcal{L}_k}\log(\epsilon_{k,l}^{-1}) + \sum\limits_{g\in\mathcal{G}}\underset{k\in\mathcal{K}_g}{\min}\log(\epsilon_{k,c}^{-1})}{\frac{1}{\eta}(\sum\limits_{k\in\mathcal{K}}\sum\limits_{l\in\mathcal{L}_k}||\mathbf{w}_{k,l}||_2^2 + \sum\limits_{g\in\mathcal{G}}||\tilde{\mathbf{w}}_{g,c}||_2^{2}) + P_{\text{cir}} }\\
\hspace{-7mm} \st
&  & \hspace{-7mm} \sum\limits_{k\in\mathcal{U}_b}\sum\limits_{l\in\mathcal{L}_k}||\mathbf{w}_{k,l}||_2^2  + \sum\limits_{g\in\mathcal{G}_b}||\tilde{\mathbf{w}}_{g,c}||_2^{2} \leq P_b,\; \forall b \in \mathcal{B} \label{eq:EEmax:reform1:MaxPC}\\
 &   & \hspace{-7mm} \underset{k\in\mathcal{K}_g}{\min}\log(\epsilon_{k,c}^{-1}) \geq \bar{R}_{g,c}, \forall g \in \mathcal{G} \label{eq:EEmax:reform1:min}
\end{eqnarray}
\end{subequations}
where the MSE expressions are given in \eqref{MSEprivate} and \eqref{MSEcommon}.
To further find a more tractable formulation, we equivalently modify \eqref{eq:EEmax:reform1} as
\begin{subequations}
\label{eq:EEmax:reform2}
\begin{eqnarray}
& \hspace{-15mm} \underset{\mathbf{w},\mathbf{u},\boldsymbol\nu}{\maxi}  & \hspace{-2mm} \frac{\sum\limits_{k\in\mathcal{K}}\sum\limits_{l\in\mathcal{L}_k}\log(\nu_{k,l}) + \sum\limits_{g\in\mathcal{G}}\underset{k\in\mathcal{K}_g}{\min}\log(\nu_{k,c})}{\frac{1}{\eta}(\sum\limits_{k\in\mathcal{K}}\sum\limits_{l\in\mathcal{L}_k}||\mathbf{w}_{k,l}||_2^2 + \sum\limits_{g\in\mathcal{G}}||\tilde{\mathbf{w}}_{g,c}||_2^{2}) + P_{\text{cir}} }\\
\hspace{-10mm} \st
&  & \hspace{-7mm} \sum\limits_{k\in\mathcal{U}_b}\sum\limits_{l\in\mathcal{L}_k}||\mathbf{w}_{k,l}||_2^2  + \sum\limits_{g\in\mathcal{G}_b}||\tilde{\mathbf{w}}_{g,c}||_2^{2} \leq P_b,\; \forall b \in \mathcal{B} \label{eq:EEmax:reform1:MaxPC}\\
 &   & \hspace{-7mm} \log(\nu_{k,c}) \geq \bar{R}_{g,c}, \forall g \in \mathcal{G}, k \in \mathcal{K}_g \label{eq:EEmax:reform2:nomin} \\
 &   & \hspace{-7mm} \epsilon_{k,l}(\mathbf{w},\mathbf{u}_{k,l}) \leq \nu_{k,l}^{-1}, \forall k \in \mathcal{K}, l \in \mathcal{L}_k \label{eq:PrivateDC} \\
 &   & \hspace{-7mm} \epsilon_{k,c}(\mathbf{w},\mathbf{u}_{k,c}) \leq \nu_{k,c}^{-1}, \forall k \in \mathcal{K} \label{eq:CommonDC}
\end{eqnarray}
\end{subequations}
where $\boldsymbol\nu \triangleq \{\{\nu_{k,l}\}_{k \in \mathcal{K}, l \in \mathcal{L}_k}, \{\nu_{k,c}\}_{k \in \mathcal{K}}\}$ are new stream specific variables denoting the inverse of the MSE of each stream at each user. In \eqref{eq:EEmax:reform2:nomin}, we have got rid of the minimum function in \eqref{eq:EEmax:reform1:min} by using the fact that all the users in the multicasting group have to satisfy the common rate constraint.
Then, to find a convex formulation for the numerator of the objective function, we replace the above formulation with
\begin{subequations}
\label{eq:EEmax:reform3}
\begin{eqnarray}
& \hspace{-15mm} \underset{\mathbf{w},\mathbf{u},\boldsymbol\nu,\mathbf{r}}{\maxi} &  \hspace{-2mm}  \frac{\sum\limits_{k\in\mathcal{K}}\sum\limits_{l\in\mathcal{L}_k}\log(\nu_{k,l}) + \sum_{g\in\mathcal{G}}r_{g,c}}{\frac{1}{\eta}(\sum\limits_{k\in\mathcal{K}}\sum\limits_{l\in\mathcal{L}_k}||\mathbf{w}_{k,l}||_2^2 + \sum\limits_{g\in\mathcal{G}}||\tilde{\mathbf{w}}_{g,c}||_2^{2}) + P_{\text{cir}} } \label{eq:concaveconvexobj}\\
\hspace{-10mm} \st
&  & \hspace{-7mm} \sum\limits_{k\in\mathcal{U}_b}\sum\limits_{l\in\mathcal{L}_k}||\mathbf{w}_{k,l}||_2^2 + \sum\limits_{g\in\mathcal{G}_b}||\tilde{\mathbf{w}}_{g,c}||_2^{2} \leq P_b,\; \forall b \in \mathcal{B} \label{eq:EEmax:reform1:MaxPC}\\
 &   & \hspace{-7mm} r_{g,c} \geq \bar{R}_{g,c}, \forall g \in \mathcal{G} \label{eq:EEmax:reform1:C1}\\
 &   & \hspace{-7mm} r_{g,c} \leq \log(\nu_{k,c}), \forall g \in \mathcal{G}, k \in \mathcal{K}_g \label{eq:EEmax:reform1:C2} \\
 &   & \hspace{-7mm} \eqref{eq:PrivateDC}, \eqref{eq:CommonDC}
\end{eqnarray}
\end{subequations}
where $r_{g,c}$ denotes the data rate of the multicasting stream for user group $g$.
The optimal receivers for the above problem are the MMSE receivers \cite{Shi:WMMSE:2011}
\begin{eqnarray}\label{MMSEprivate}
\mathbf{u}_{k,l} = (\sum_{i\in\mathcal{K}}\sum_{j\in\mathcal{L}_i}\mathbf{H}_{b_i,k}\mathbf{w}_{i,j}\mathbf{w}_{i,j}\herm\mathbf{H}_{b_i,k}\herm \nonumber\\
+ \sum_{u\in\mathcal{G}}\mathbf{H}_{b_u,k}\tilde{\mathbf{w}}_{u,c}\tilde{\mathbf{w}}_{u,c}\herm\mathbf{H}_{b_u,k}\herm + N_0\mathbf{I})^{-1}\mathbf{H}_{b_k,k}\mathbf{w}_{k,l}
\end{eqnarray}
\begin{eqnarray}\label{MMSEcommon}
\mathbf{u}_{k,c} = (\sum_{j\in\mathcal{K}}\sum_{i\in\mathcal{L}_j}\mathbf{H}_{b_i,k}\mathbf{w}_{i,j}\mathbf{w}_{i,j}\herm\mathbf{H}_{b_i,k}\herm \nonumber \\
+ \sum_{u\in\mathcal{G}}\mathbf{H}_{b_u,k}\tilde{\mathbf{w}}_{u,c}\tilde{\mathbf{w}}_{u,c}\herm\mathbf{H}_{b_u,k}\herm + N_0\mathbf{I})^{-1}\mathbf{H}_{b_k,k}\tilde{\mathbf{w}}_{g,c}.
\end{eqnarray}
If the receivers are fixed, the objective function \eqref{eq:concaveconvexobj} is a concave-convex fractional function and all the other constraints are convex except \eqref{eq:PrivateDC} and \eqref{eq:CommonDC} which involve difference of convex functions. We can write the linear lower approximations of \eqref{eq:PrivateDC} as
\begin{eqnarray}\label{approx}
\nu_{k,l}^{-1} \geq (\nu_{k,l}^{(n)})^{-1} - (\nu_{k,l}^{(n)})^{-2}(\nu_{k,l} - \nu_{k,l}^{(n)}) \triangleq \Xi_{k,l}^{(n)}(\nu_{k,l})
\end{eqnarray}
and the approximation for \eqref{eq:CommonDC} equivalently.
With the approximations, we get the following problem
\begin{subequations}
\label{eq:EEmax:reform4}
\begin{eqnarray}
& \hspace{-15mm} \underset{\mathbf{w},\boldsymbol\nu,\mathbf{r}}{\maxi} &  \hspace{-2mm}  \frac{\sum\limits_{k\in\mathcal{K}}\sum\limits_{l\in\mathcal{L}_k}\log(\nu_{k,l}) + \sum_{g\in\mathcal{G}}r_{g,c}}{\frac{1}{\eta}(\sum\limits_{k\in\mathcal{K}}\sum\limits_{l\in\mathcal{L}_k}||\mathbf{w}_{k,l}||_2^2 + \sum\limits_{g\in\mathcal{G}}||\tilde{\mathbf{w}}_{g,c}||_2^{2}) + P_{\text{cir}} }\\
\hspace{-10mm} \st
 &   & \hspace{0mm} \epsilon_{k,l}(\mathbf{w},\mathbf{u}_{k,l}) \leq \Xi_{k,l}^{(n)}(\nu_{k,l}), \forall k \in \mathcal{K}, l \in \mathcal{L}_k \\
 &   & \hspace{0mm} \epsilon_{k,c}(\mathbf{w},\mathbf{u}_{k,c}) \leq \Xi_{k,c}^{(n)}(\nu_{k,c}), \forall k \in \mathcal{K}\\
 &   & \eqref{eq:EEmax:reform1:MaxPC}, \eqref{eq:EEmax:reform1:C1}, \eqref{eq:EEmax:reform1:C2}.
\end{eqnarray}
\end{subequations}
At this point, we note that the problem is a concave-convex fractional program which can be transformed to a convex one with the Charnes-Cooper transformation \cite{Schaible-76}. Thus, solving \eqref{eq:EEmax:reform4} is equivalent to the following convex problem
\begin{subequations}
\label{eq:EEmax:reform5}
\begin{eqnarray}
& \hspace{-10mm} \underset{\bar{\mathbf{w}},\bar{\boldsymbol\nu},\bar{\mathbf{r}},\phi}{\maxi}  & \hspace{-2mm} \sum\limits_{k\in\mathcal{K}}\sum\limits_{l\in\mathcal{L}_k}\phi\log(\frac{\bar{\nu}_{k,l}}{\phi}) + \sum_{g\in\mathcal{G}}\bar{r}_{g,c}\label{eq:EEmax:reform4:obj}\\
\hspace{-10mm} \st
&  & \hspace{-8mm} \sum\limits_{k\in\mathcal{U}_b}\sum\limits_{l\in\mathcal{L}_k}||\bar{\mathbf{w}}_{k,l}||_2^2 + \sum\limits_{g\in\mathcal{G}_b}||\bar{\tilde{\mathbf{w}}}_{g,c}||_2^2 \leq \phi^2 P_b,\; \forall b \in \mathcal{B} \label{eq:EEmax:reform4:MaxPC}\\
 &   & \hspace{-8mm} \bar{r}_{g,c} \geq \phi\bar{R}_{g,c}, \forall g \in \mathcal{G}\\
 &   & \hspace{-8mm}  \phi\epsilon_{k,l}(\frac{\bar{\mathbf{w}}}{\phi},\mathbf{u}_{k,l}) \leq \Xi_{k,l}^{(n)}(\frac{\bar{\nu}_{k,l}}{\phi}), \forall k \in \mathcal{K}, l \in \mathcal{L}_k \\
 &   & \hspace{-8mm} \phi\epsilon_{k,c}(\frac{\bar{\mathbf{w}}}{\phi},\mathbf{u}_{k,c}) \leq \Xi_{k,c}^{(n)}(\frac{\bar{\nu}_{k,c}}{\phi}), \forall k \in \mathcal{K}\\
 &   & \hspace{-8mm} \bar{r}_{g,c} \leq \phi\log(\frac{\bar{\nu}_{k,c}}{\phi}), \forall g \in \mathcal{G}, k \in \mathcal{K}_g\\
 &   & \hspace{-8mm} \frac{1}{\eta}(\sum\limits_{k\in\mathcal{K}}\sum\limits_{l\in\mathcal{L}_k}||\bar{\mathbf{w}}_{k,l}||_2^2 + \sum\limits_{g\in\mathcal{G}}||\bar{\tilde{\mathbf{w}}}_{g,c}||_2^2) + \phi^2 P_{\text{cir}} \leq \phi.
\end{eqnarray}
\end{subequations}
After solving \eqref{eq:EEmax:reform5}, the optimal solutions for the original fractional program \eqref{eq:EEmax:reform4} can be found as $\mathbf{w}^* = \bar{\mathbf{w}}^*/\phi^*, \nu_{k,c}^* = \frac{\bar{\nu}_{k,c}^*}{\phi^*}, \nu_{k,l}^* = \frac{\bar{\nu}_{k,l}^*}{\phi^*},r_{g,c}^* = \frac{\bar{r}_{g,c}^*}{\phi^*}$, where $\bar{\mathbf{w}}^*,\phi^*, \bar{\nu}_{k,c}^*,\bar{\nu}_{k,l}^*,\bar{r}_{g,c}^*$ are the optimal variables of \eqref{eq:EEmax:reform5}.
The method to solve the EEmax problem \eqref{EEmax} is presented in Algorithm \ref{algo:iterative}. The idea is to use alternating optimization between the transmit beamformers and receivers, so that for fixed receivers, successive convex approximation is used to solve \eqref{eq:EEmax:reform5} iteratively. However, to improve the convergence speed, the receivers and linearization point (i.e, $\Xi_{k,l}^{(n)}(\frac{\bar{\nu}_{k,l}}{\phi})$ and $\Xi_{k,c}^{(n)}(\frac{\bar{\nu}_{k,c}}{\phi})$) are both updated every time after solving \eqref{eq:EEmax:reform5}. This also guarantees monotonicity of the objective function \eqref{eq:EEmax:reform4:obj} \cite{Venkatraman-16, Tervo-16Arxiv}. We note that a solution feasible to \eqref{eq:EEmax:reform4} is also feasible to the original problem due to the approximation in \eqref{approx}. That is, the feasible set of \eqref{eq:EEmax:reform4} is an inner convex set of the original problem and will be updated as the iterative procedure progresses. Thus, the solution returned by Algorithm \ref{algo:iterative} is not guaranteed to be globally optimal to the original nonconvex problem in \eqref{EEmax}. The expectation is that the produced solution is sufficiently good as the approximated feasible set is improved after each iteration. This is the main principle of SCA framework to deal with nonconvex optimization problems. For a more detailed convergence analysis, it is possible to modify \cite[Appendix A]{Venkatraman-16} to see the convergence properties of the algorithm.

\begin{algorithm}[t]
\begin{algorithmic}[1] \caption{Proposed joint multicasting and unicasting beamforming design.}
\label{algo:iterative} \renewcommand{\algorithmicrequire}{\textbf{Initialization:}}
\REQUIRE Set $n=0$, and generate feasible initial $\boldsymbol\nu^{(0)}$.
\REPEAT
\STATE Solve \eqref{eq:EEmax:reform5} with $\boldsymbol\nu^{(n)}$
and denote optimal values as $\bar{\boldsymbol\nu}^*$.
\STATE Update $\boldsymbol\nu^{(n+1)} = \bar{\boldsymbol\nu}^*/\phi$ and $\Xi_{k,l}^{(n+1)}(\nu_{k,l}), \forall k \in \mathcal{K}, l \in \mathcal{L}_k, \Xi_{k,c}^{(n+1)}(\nu_{k,c}), \forall k \in \mathcal{K}$.\label{algo:update}
\STATE Update MMSE receivers according to \eqref{MMSEprivate} and \eqref{MMSEcommon}.
\STATE $n:=n+1$.
\UNTIL desired accuracy level
\renewcommand{\algorithmicrequire}{\textbf{Output:}} \REQUIRE $\tilde{\mathbf{w}}_{g,c}^* = \tfrac{\bar{\tilde{\mathbf{w}}}_{g,c}^*}{\phi^*}, \forall g \in \mathcal{G}, \mathbf{w}_{k,l}^* = \tfrac{\bar{\mathbf{w}}_{k,l}^*}{\phi^*}, \forall k \in \mathcal{K}, l \in \mathcal{L}_k, \mathbf{u}_{k,c}^*, \forall k \in \mathcal{K}, \mathbf{u}_{k,l}^*, \forall k \in \mathcal{K}, l \in \mathcal{L}_k$
\end{algorithmic}
\end{algorithm}

\section{Numerical Results}
\label{sec:NumericalResults}

We evaluate the performance for a quasistatic frequency flat Rayleigh fading channel model with $B=2$ BSs. Each BS serves $K/2$ users which are randomly divided into $L/2$ groups of users of equal size, i,e, $K$ is the total number of users and $L$ is the total number of groups in the network. The distance from the BS to all the users is 250m, and we use the cell separation parameter $\mu=3$ [dB] (cf. \cite{Komulainen-13} for further details on this parameter) to control the average interference between the cells. As a result, the total path loss from BS $b$ to user of a neighboring cell is calculated as $\gamma=35+30\log_{10}(250)+\mu$ [dB], where the first term is the distance dependent path loss, and the average path loss from BS to its own users is $\gamma=35+30\log_{10}(250)$. The bandwidth is set to $W=20$ MHz and noise power $N_0=-125$ dBW. The number of antennas is $N_b = N$ for all $b$, i.e., $N$ is the number of antennas at each BS, and $M_k=M, \forall k \in \mathcal{K}$ is the number of receiver antennas. The power consumption parameters are $\eta=0.35, P_{\text{max}}=3$ dBW, $P_{\text{0,BS}}=1$ Watts, $P_{\text{0,UE}}=0.2$ Watts, $P_{\text{RF,BS}}=0.4$ Watts, $P_{\text{RF,UE}}=0.2$ Watts. Also, equal rate targets are assumed for each group, i.e., $\bar{R}_{g,c}=\bar{R}, \forall g \in \mathcal{G}$, and the other simulation parameters are given in the figures.

\begin{figure}[t]
\centering
  \includegraphics[width=0.86\columnwidth]{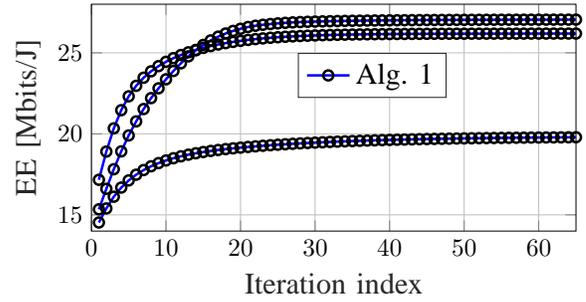}
  \caption{Convergence of Alg. 1 for three random channel realizations with $N=4,M=2,K=8,L=4,\bar{R}=72.14 \ \text{Mbits/s}$.}
  \label{fig: Convergence}
\end{figure}

\begin{figure}[t]
\centering
  \includegraphics[width=0.86\columnwidth]{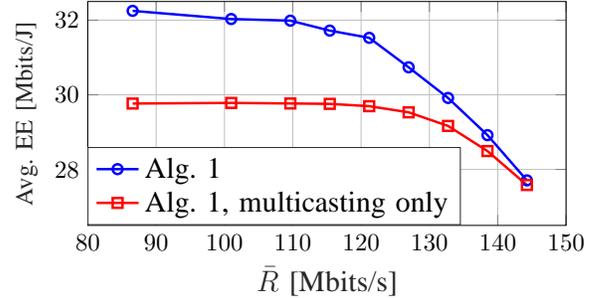}
  \caption{Energy efficiency versus the minimum common stream rate constraint with $N=8,M=2,K=8,L=4$.}
  \label{fig: EEvsRcmin}
\end{figure}

\begin{figure}[t]
\centering
  \includegraphics[width=0.86\columnwidth]{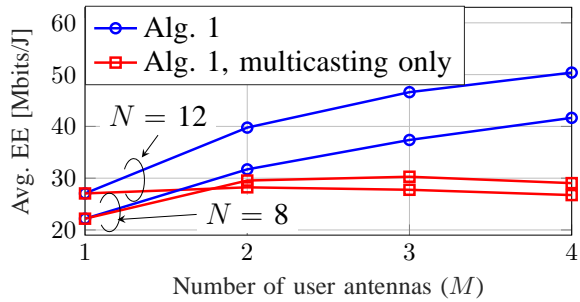}
  \caption{Energy efficiency versus the number of antennas per user with $K=8,L=4,\bar{R}=115.42 \ \text{Mbits/s}$.}
  \label{fig: EEvsM}
\end{figure}

\begin{figure}[t]
\centering
  \includegraphics[width=0.86\columnwidth]{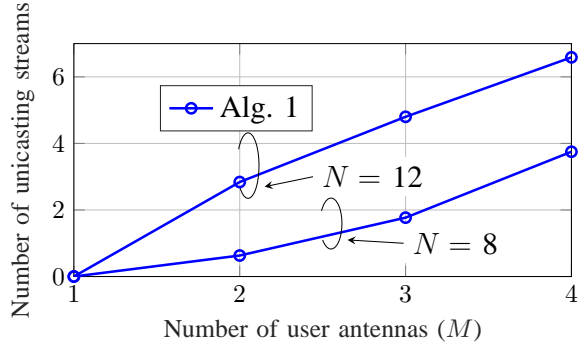}
  \caption{The average number of transmitted unicasting streams versus the number of user antennas with $K=8,L=4,\bar{R}=115.42 \ \text{Mbits/s}$.}
  \label{fig: nStreams}
\end{figure}

Fig. \ref{fig: Convergence} illustrates the convergence of Algorithm 1 for three different channel realizations. We can observe the monotonicity of the objective function,
and that most of the gains are achieved during the first 10 iterations.

Fig. \ref{fig: EEvsRcmin} plots the energy efficiency versus minimum rate constraint for the common streams. It is observed that energy efficiency gains can be achieved specifically when the rate constraints for the multicasting streams are lower. Obviously, when the rate constraints are large, all the resources are consumed to satisfy the multicasting constraints.

Fig. \ref{fig: EEvsM} demonstrates the effect of user antennas on the average energy efficiency for $N=8$ and $N=12$. We can see that if only multicasting information is transmitted, the network energy efficiency begins to decrease when the users have more antennas. The reason is that the antennas are only used to harvest diversity gains, while the potential of transmitting additional unicasting streams is not exploited. Thus, the power consumption increase due to the additional number of antennas overwhelms the achievable diversity gain. However, when the extra degrees of freedom provided by the user antennas are used to transmit additional streams, the energy efficiency improvement is significant. The number of transmitted unicasting streams to achieve the highest energy efficiency is shown in Fig. \ref{fig: nStreams}. It is observed that when the number of user antennas increases, it is energy-efficient to transmit more unicasting streams. It is worth mentioning that when $M=1$, the proposed algorithm gives the same average result as the algorithm in \cite{Tervo-17} for fixed antenna sets. However, the algorithm in \cite{Tervo-17} cannot solve the problem when $M>1$.

\section{Conclusions}
\label{sec:Conclusions}
This paper has studied energy efficiency optimization for multi-cell multigroup coordinated joint transmit and receive beamforming, where the users can receive unicasting data in addition to the common multicasting information. The resulting non-convex fractional program was tackled by successive convex approximation, where the transmit beamformers, linearization point, and receive beamformers are sequentially optimized. The numerical results have illustrated that it is energy-efficient to use the additional transmission resources for unicasting services when the users have multiple receive antennas, and the gains increase with the number of antennas. The network energy efficiency decreases with the number of user antennas, if only multicasting transmission is used.

\section{Acknowledgements}
This work was supported by Infotech Oulu Doctoral Program and the Academy of Finland under project WiConIE. It was also supported by a research grant from Science Foundation Ireland and is co-funded by the European Regional Development Fund under Grant 13/RC/2077, projects FNR SEMIGOD, SATSENT, INWIPNET and H2020 SANSA. The first author has been supported by KAUTE Foundation, the Finnish Foundation for Technology Promotion, HPY Research Foundation, Walter Ahlström Foundation, Tauno Tönning Foundation, Seppo Säynäjäkankaan Tiedesäätiö and Nokia Foundation. 

\bibliographystyle{IEEEtran}

\end{document}